\documentclass[a4paper,11pt]{article}
\usepackage{pos}
\usepackage{color,soul}

\title{VERITAS follow-up observation of the blazar TXS 0506+056 }
 \ShortTitle{VERITAS follow-up observation of the blazar TXS 0506+056}

\author*[a]{Weidong Jin}
\author[]{on behalf of the VERITAS Collaboration}

\affiliation[a]{Department of Physics and Astronomy, University of Alabama, Tuscaloosa, AL 35487-0324, USA}

 \forColl{}  

\author[a]{RileyAnne Sharpe}

\emailAdd{wjin4@crimson.ua.edu}
\emailAdd{rhsharpe@crimson.ua.edu}

\abstract{The gamma-ray blazar TXS 0506+056 was found with an enhanced gamma-ray emission state in spatial and temporal coincidence with the IceCube high energy neutrino event IC170922A. This is the most significant association by far between a high-energy neutrino event and a blazar in a flaring state. Studying the time evolution and spectral behavior of the blazar emission may help in identifying the sources of the diffuse neutrino flux observed by IceCube and the origin of energetic cosmic rays. TXS 0506+056 was detected by the VERITAS gamma-ray observatory with a significance of 5.8 standard deviations above 110 GeV in a 35 hour data set collected between September 23, 2017 and February 6, 2018. Here we will present results from recent VERITAS observations and an associated multiwavelength campaign, collected between October 10, 2018 to March 1, 2021. A relatively quiet very high energy gamma-ray emission state was observed during this time period, and flux upper limits are used to constrain the potential variability of this blazar.}

\FullConference{37$^{\rm{th}}$ International Cosmic Ray Conference (ICRC 2021)\\
		July 12th -- 23rd, 2021\\
		Online -- Berlin, Germany}

\begin{document}
\maketitle

\section{Introduction}

On September 22, 2017, the IceCube Collaboration reported the detection of a high-energy muon neutrino (290 TeV) event, labeled IceCube-170922A, of potential astrophysical origin. This event was sent out as an alert \citep{GCN21916} and prompted a multiwavelength campaign from the radio to very-high-energy (VHE, E $>$ 100 GeV) gamma-ray band. The location of the neutrino was consistent with that of the blazar TXS 0506+056 (z = 0.3365 $\pm$ 0.0010) which at the time was observed by \emph{Fermi}-LAT to be flaring in gamma rays. The correlation of the neutrino detection with the gamma-ray flare of TXS 0506+056 is statistically significant at the 3$\sigma$ level \citep{eaat1378}. Follow-up observations led to the first detection of this blazar in the VHE emission region by the MAGIC telescopes \citep{ansoldi2018blazar}. An initial observation taken by VERITAS did not yield a detection, but a follow-up observation taken between September 23, 2017 and February 6, 2018 (MJD 58019-58155) yielded a detection of the source above 110 GeV with a statistical significance of 5.8$\sigma$ based on a 35-hr data set \citep{Abeysekara_2018}. 

Although no significant excess of cumulative neutrino emission was found by searching gamma-ray blazars detected by 2nd \emph{Fermi}-LAT AGN catalog (2LAC), and the maximum contribution from blazars in 2LAC to the all-sky astrophysical neutrino flux is constrained to be no more than $\sim$27\% \citep{Aartsen_2017_fermi}, this did not rule out that individual blazars could be potential neutrino counterparts. Since 2017, TXS 0506+056 remains the most significant correlation between a blazar and a high-energy neutrino. More recently, other blazars such as 4FGL J0658.6+0636 (more details in this conference in \citep{201114A}) has also been identified in the 90\% uncertainty region of well-reconstructed IceCube neutrino events. Thus TXS 0506+056 plays a key role in establishing connections between high-energy neutrinos and astrophysical sources. Studying the time evolution and spectral behavior of this blazar may help in identifying the sources of the diffuse neutrino flux observed by IceCube and the origin of energetic cosmic rays.

\section{VERITAS observation of blazar TXS 0506+056 from 2018 to 2021}
VERITAS \cite{Holder2006} consists of an array of four 12-m imaging atmospheric Cherenkov telescopes located at the Fred Lawrence Whipple Observatory (FLWO) in southern Arizona, USA (31$^{\circ}$ 40' N, 110$^{\circ}$ 57' W, 1.3 km a.s.l.). Each telescope equipped with a camera containing 499 photomultiplier tubes covering a field of view of 3.5$^{\circ}$. VERITAS has maximum sensitivity in the range of 100 GeV to 30 TeV. The angular resolution of VERITAS is $\sim0.1^{\circ}$ at 1 TeV (for 68\% containment) and the energy resolution is 15-25\% at the same energy. In its current configuration, VERITAS can detect a source ($5\sigma$ significance) with a flux of 1\% of the steady flux from the Crab Nebula in less than 25 hours \cite{2015ICRC...34..771P}. VERITAS data were analyzed using standard analysis tools~\cite{2017ICRC...35..747M, 2008ICRC....3.1385C, 2008ICRC....3.1325D}.

\begin{figure}[ht!]
\begin{center}
    \includegraphics[width=0.5\textwidth]{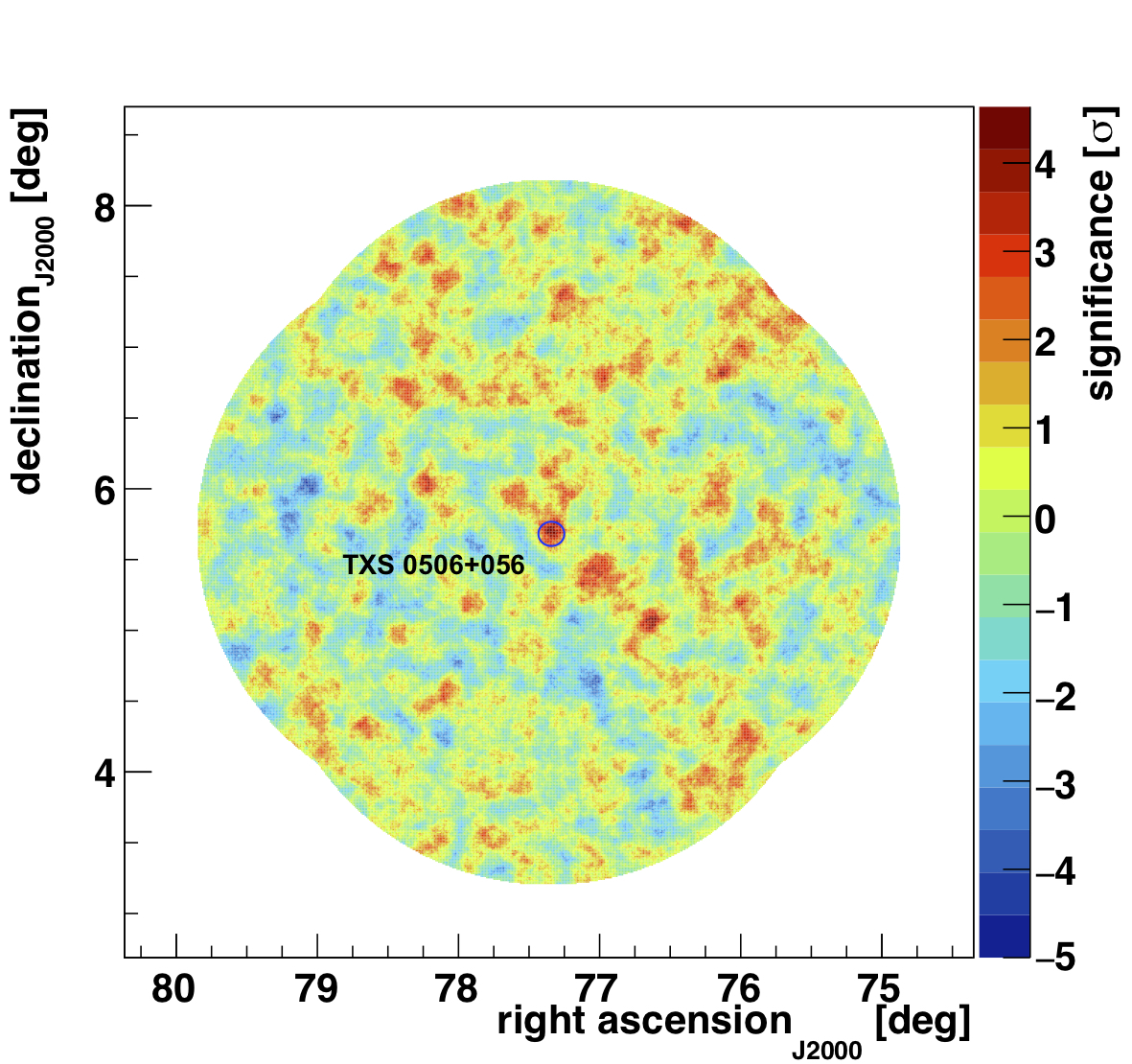}
    \caption{VERITAS significance sky map with TXS 0506+056 in the center.}
    \label{fig:skymap}
\end{center}
\end{figure}

VERITAS has implemented a wide-ranging neutrino follow-up program to search for VHE gamma rays associated with IceCube neutrino events~\citep{santander2019recent}, including  follow-up of real-time IceCube neutrino alerts (more details at this conference in \citep{ICRC2021_NToO}) and long-term monitoring of the candidate neutrino blazar TXS 0506+056. VERITAS collected 61 hours of quality-selected observations of TXS 0506+056 from from Oct 10, 2018 (MJD 58401) to March 7, 2021 (MJD 59280) with an average zenith angle of 28.8$^{\circ}$. Observations were performed using the standard “wobble” observation mode~\citep{Fomin1994} with a 0.5$^{\circ}$ offset in each of four cardinal directions. The analysis yielded a detection of the source above 190 GeV with a statistical significance of 3.4$\sigma$. The integral flux above an energy threshold of 190 GeV is $(1.34\pm0.40)\times10^{-12}$ cm$^{-2}$ s$^{-1}$, which corresponds to $(0.52\pm0.16)\%$ of the Crab Nebula flux (C.U.) under the same energy threshold \citep{Hillas1998,Meagher2015ICRC}. Fig. \ref{fig:skymap} shows a significance sky map with TXS 0506+056 in the center. A light curve of the integral flux above an energy threshold of 190 GeV is shown in Fig. \ref{fig:multi_lc}. Each point represents the average flux level of the source with one month per bin. The vertical gray lines represent individual observations. A 95$\%$ confidence level upper limit is calculated when significance is less than 2$\sigma$. For the upper limit calculation, this analysis used the best fitting spectral index ($\Gamma$=4.8) obtained from VERITAS's follow-up observation of TXS 0506+056 from year 2017 to 2018 \citep{Abeysekara_2018}. A relatively quiet VHE gamma-ray emission state was observed.

\section{Multiwavelength observations}
\begin{figure}[ht!]
\begin{center}
    \includegraphics[width=0.9\textwidth]{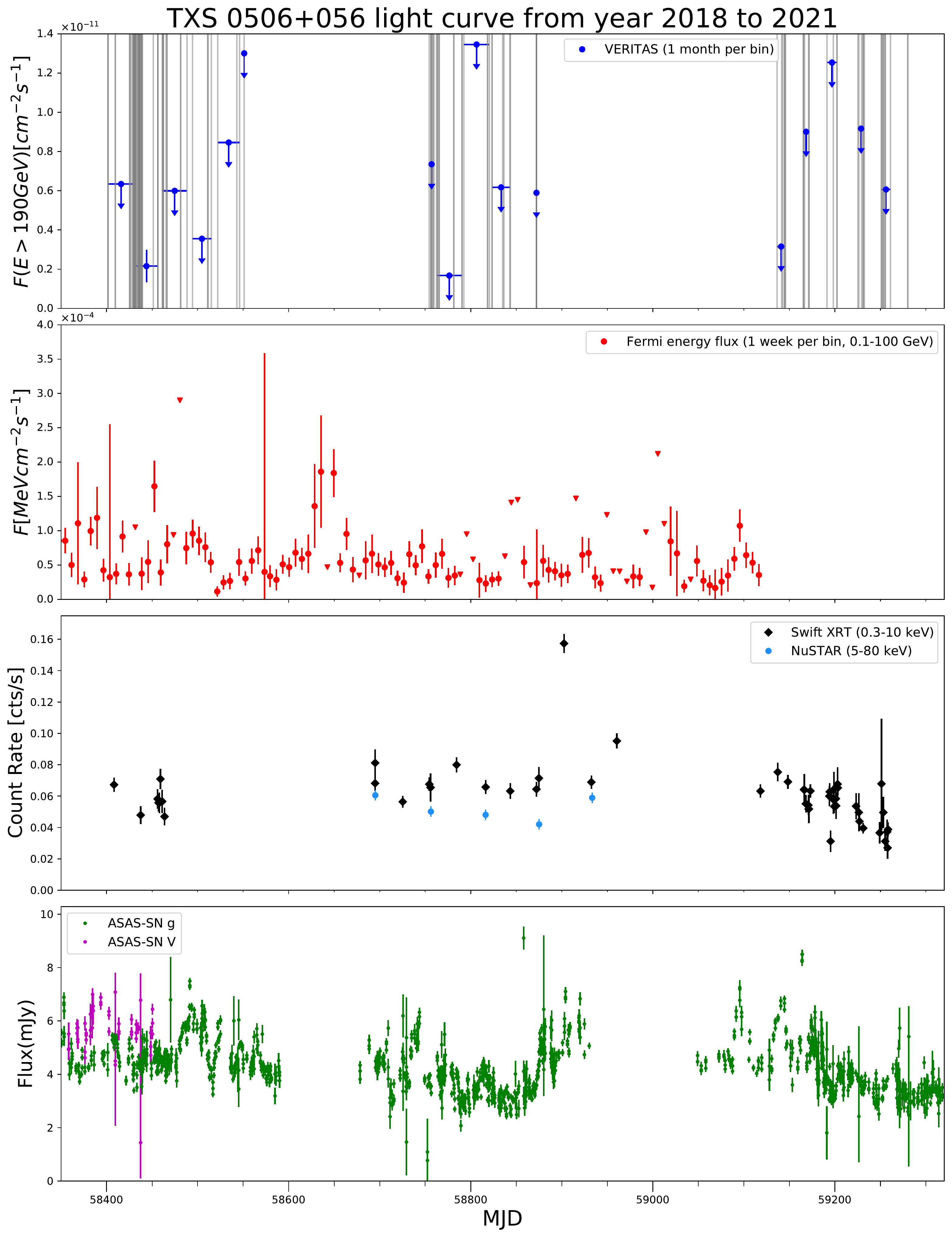}
    \caption{Multiwavelength light curves of blazar TXS 0506+056 from Oct 10, 2018 to March 7, 2021. The data set is composed of VHE gamma-ray data from VERITAS, high-energy gamma-ray data from \emph{Fermi}-LAT LCR, X-ray data from \emph{Swift} and NuSTAR, and optical data (not corrected for extinction) from ASAS-SN Sky Patrol.}
    \label{fig:multi_lc}
\end{center}
\end{figure}

The multiwavelength light curves are composed of high energy gamma-ray data from the \emph{Fermi}-LAT light curve repository (LCR)\footnote{\url{https://fermi.gsfc.nasa.gov/ssc/data/access/lat/LightCurveRepository/}}, X-ray data from the Neil Gehrels \emph{Swift} Observatory X-ray telescope (XRT)\footnote{\url{https://www.swift.ac.uk/user_objects/}} and Nuclear Spectroscopic Telescope Array (NuSTAR), as well as optical data from the All-Sky Automated Survey for Supernovae (ASAS-SN) Sky Patrol\footnote{\url{https://asas-sn.osu.edu/}} (Fig. \ref{fig:multi_lc}). The LCR is a public library of light curves for sources in the 10 year Fermi LAT point source (4FGL) catalog \citep{Abdollahi2020}. The data are presented on time scales of weeks and photon energy between 100 MeV and 100 GeV. The observation made by ASAS-SN Sky Patrol are available in two optical band: the \emph{V} (original two stations) or \emph{g} (three new stations) band filters and three dithered 90s exposures \citep{2014Shappee,2017Kochanek}.

NuSTAR and \emph{Swift} observations were collected during a monitoring campaign of TXS 0506+056 to support the VERITAS observations in search for evidence of hadronic emission. The campaign, organized during NuSTAR GO Cycle 5 between MJD 58550 and 58950, consisted of 4 $\times$ 20 ks NuSTAR exposures separated by $\sim 60$ days to sample similar time scales as the candidate neutrino ``flare'' of 2014-2015 and was supplemented by 10 $\times$ 3 ks \emph{Swift} exposures interleaved with the NuSTAR epochs. The epochs were selected to coincide with the middle of the VERITAS dark observation period so that quasi-simultaneous observations were possible. The NuSTAR and \emph{Swift}-XRT observations were reduced using the \texttt{nupipeline} and \texttt{xrtpipeline} routines, respectively, available in HEASoft 6.28. Background-subtracted count rate light curves were derived from the reduced data in two energy ranges: 0.3 - 10 keV for \emph{Swift}-XRT and 5-80 keV for NuSTAR. The light curves shown in Fig.~\ref{fig:multi_lc} show no indication of strong flaring in X-rays with the exception of an elevated state observed by \emph{Swift} on MJD 58902 with no apparent counterpart in \emph{Fermi}. The flux level that corresponds to this rate ($\sim 4 \times 10^{-12}$ erg cm$^{-2}$ s$^{-1}$) did not meet the pre-defined criteria to trigger an additional NuSTAR exposure during the monitoring program and no coincident VERITAS observations were possible at the time as it happened outside the VERITAS dark observing period. The detailed analysis of the X-ray data is ongoing. 

\section{Summary and outlook}
VERITAS's previous follow-up observations of TXS 0506+056 during 2017/2018 season gave a flux of $(1.79\pm0.56)\times10^{-12}$ cm$^{-2}$ s$^{-1}$, corresponding to $(0.70\pm0.22)\%$ of C.U. (the analysis was redone with energy threshold corrected to 190 GeV and updated instrument response functions) \citep{Abeysekara_2018}. TXS 0506+056 is found to be in a quiet state from 2018 to 2021 and shows a consistent flux level $((0.52\pm0.16)\%$ of C.U.) compared to previous VERITAS follow-up observations. MAGIC detected enhanced VHE gamma-ray emissions on Dec $1^{st}$ (MJD 58453) and $3^{rd}$ 2018 (MJD 58455) and its flux was reported as comparable with the flare detected by MAGIC in October 2017 \citep{satalecka2019magic}. This enhanced emission was not detected by VERITAS because VERITAS didn't have observation on the first date and the data quality is bad on the second date due to poor weather condition. In addition to that, no extended VHE gamma-ray emissions were detected by VERITAS, consistent with a low state of the source during this period.  Clear variability is observed in optical, X-ray and high-energy gamma-ray bands. Quantitative analysis will be presented in an upcoming publication.

\section*{Acknowledgements}
This research is supported by grants from the U.S. Department of Energy Office of Science, the U.S. National Science Foundation and the Smithsonian Institution, by NSERC in Canada, and by the Helmholtz Association in Germany. This research used resources provided by the Open Science Grid, which is supported by the National Science Foundation and the U.S. Department of Energy's Office of Science, and resources of the National Energy Research Scientific Computing Center (NERSC), a U.S. Department of Energy Office of Science User Facility operated under Contract No. DE-AC02-05CH11231. We acknowledge the excellent work of the technical support staff at the Fred Lawrence Whipple Observatory and at the collaborating institutions in the construction and operation of the instrument.

This research was partially supported by NASA grant NuSTAR GO-5277. This research has made use of data obtained with NuSTAR, a project led by Caltech, funded by NASA and managed by NASA/JPL, and has utilized the NUSTARDAS software package, jointly developed by the ASDC (Italy) and Caltech (USA).

\bibliographystyle{JHEP}
\bibliography{references.bib}

\clearpage
\section*{Full Authors List: \Coll\ VERITAS Collaboration}


\scriptsize
\noindent
C.~B.~Adams$^{1}$,
A.~Archer$^{2}$,
W.~Benbow$^{3}$,
A.~Brill$^{1}$,
J.~H.~Buckley$^{4}$,
M.~Capasso$^{5}$,
J.~L.~Christiansen$^{6}$,
A.~J.~Chromey$^{7}$, 
M.~Errando$^{4}$,
A.~Falcone$^{8}$,
K.~A.~Farrell$^{9}$,
Q.~Feng$^{5}$,
G.~M.~Foote$^{10}$,
L.~Fortson$^{11}$,
A.~Furniss$^{12}$,
A.~Gent$^{13}$,
G.~H.~Gillanders$^{14}$,
C.~Giuri$^{15}$,
O.~Gueta$^{15}$,
D.~Hanna$^{16}$,
O.~Hervet$^{17}$,
J.~Holder$^{10}$,
B.~Hona$^{18}$,
T.~B.~Humensky$^{1}$,
W.~Jin$^{19}$,
P.~Kaaret$^{20}$,
M.~Kertzman$^{2}$,
T.~K.~Kleiner$^{15}$,
S.~Kumar$^{16}$,
M.~J.~Lang$^{14}$,
M.~Lundy$^{16}$,
G.~Maier$^{15}$,
C.~E~McGrath$^{9}$,
P.~Moriarty$^{14}$,
R.~Mukherjee$^{5}$,
D.~Nieto$^{21}$,
M.~Nievas-Rosillo$^{15}$,
S.~O'Brien$^{16}$,
R.~A.~Ong$^{22}$,
A.~N.~Otte$^{13}$,
S.~R. Patel$^{15}$,
S.~Patel$^{20}$,
K.~Pfrang$^{15}$,
M.~Pohl$^{23,15}$,
R.~R.~Prado$^{15}$,
E.~Pueschel$^{15}$,
J.~Quinn$^{9}$,
K.~Ragan$^{16}$,
P.~T.~Reynolds$^{24}$,
D.~Ribeiro$^{1}$,
E.~Roache$^{3}$,
J.~L.~Ryan$^{22}$,
I.~Sadeh$^{15}$,
M.~Santander$^{19}$,
G.~H.~Sembroski$^{25}$,
R.~Shang$^{22}$,
D.~Tak$^{15}$,
V.~V.~Vassiliev$^{22}$,
A.~Weinstein$^{7}$,
D.~A.~Williams$^{17}$,
and 
T.~J.~Williamson$^{10}$\\

\noindent
$^1${Physics Department, Columbia University, New York, NY 10027, USA}
$^{2}${Department of Physics and Astronomy, DePauw University, Greencastle, IN 46135-0037, USA}
$^3${Center for Astrophysics $|$ Harvard \& Smithsonian, Cambridge, MA 02138, USA}
$^4${Department of Physics, Washington University, St. Louis, MO 63130, USA}
$^5${Department of Physics and Astronomy, Barnard College, Columbia University, NY 10027, USA}
$^6${Physics Department, California Polytechnic State University, San Luis Obispo, CA 94307, USA} 
$^7${Department of Physics and Astronomy, Iowa State University, Ames, IA 50011, USA}
$^8${Department of Astronomy and Astrophysics, 525 Davey Lab, Pennsylvania State University, University Park, PA 16802, USA}
$^9${School of Physics, University College Dublin, Belfield, Dublin 4, Ireland}
$^{10}${Department of Physics and Astronomy and the Bartol Research Institute, University of Delaware, Newark, DE 19716, USA}
$^{11}${School of Physics and Astronomy, University of Minnesota, Minneapolis, MN 55455, USA}
$^{12}${Department of Physics, California State University - East Bay, Hayward, CA 94542, USA}
$^{13}${School of Physics and Center for Relativistic Astrophysics, Georgia Institute of Technology, 837 State Street NW, Atlanta, GA 30332-0430}
$^{14}${School of Physics, National University of Ireland Galway, University Road, Galway, Ireland}
$^{15}${DESY, Platanenallee 6, 15738 Zeuthen, Germany}
$^{16}${Physics Department, McGill University, Montreal, QC H3A 2T8, Canada}
$^{17}${Santa Cruz Institute for Particle Physics and Department of Physics, University of California, Santa Cruz, CA 95064, USA}
$^{18}${Department of Physics and Astronomy, University of Utah, Salt Lake City, UT 84112, USA}
$^{19}${Department of Physics and Astronomy, University of Alabama, Tuscaloosa, AL 35487, USA}
$^{20}${Department of Physics and Astronomy, University of Iowa, Van Allen Hall, Iowa City, IA 52242, USA}
$^{21}${Institute of Particle and Cosmos Physics, Universidad Complutense de Madrid, 28040 Madrid, Spain}
$^{22}${Department of Physics and Astronomy, University of California, Los Angeles, CA 90095, USA}
$^{23}${Institute of Physics and Astronomy, University of Potsdam, 14476 Potsdam-Golm, Germany}
$^{24}${Department of Physical Sciences, Munster Technological University, Bishopstown, Cork, T12 P928, Ireland}
$^{25}${Department of Physics and Astronomy, Purdue University, West Lafayette, IN 47907, USA}

\end{document}